\documentclass[conference]{IEEEtran}
\IEEEoverridecommandlockouts
\usepackage{amsmath,amssymb,amsfonts,mathtools}
\usepackage{bm}
\usepackage{graphicx}
\usepackage{booktabs}
\usepackage{multirow}
\usepackage{algorithm}
\usepackage{algpseudocode}
\usepackage{siunitx}
\usepackage{xcolor}
\usepackage{enumitem}
\usepackage{url}
\usepackage[normalem]{ulem}
\usepackage{cite}
\usepackage{subfigure}
\usepackage{array}
\usepackage{makecell}
\usepackage{dsfont}
\usepackage{color,soul}

\def\BibTeX{{\rm B\kern-.05em{\sc i\kern-.025em b}\kern-.08em
    T\kern-.1667em\lower.7ex\hbox{E}\kern-.125emX}}
\usepackage{balance}

\begin{document}

% ====== Title & authors ======
\title{Tube-Structured Incremental Semantic HARQ for Generative Video Receivers}
% Tube-Structured Semantic HARQ in Generative Video Receivers
% Tube-Structured Requests for Incremental Semantic HARQ in Generative Video Receivers
% {Incremental Semantic HARQ for Generative Video Receivers with Tube-Structured Requests}

% \author{
%     \IEEEauthorblockN{
%     Xuesong~Wang\IEEEauthorrefmark{1}\IEEEauthorrefmark{4}, Xinyan~Xie\IEEEauthorrefmark{2},
%     Runxin~Zhang\IEEEauthorrefmark{3}
%     \\
%     {wangxuesong@cuhk.edu.cn}\IEEEauthorrefmark{4}
%     }
%     \IEEEauthorblockA{\IEEEauthorrefmark{1}The Chinese University of Hong Kong, Shenzhen, 518172, China.}
%     \IEEEauthorblockA{\IEEEauthorrefmark{2}College of Smart Materials and Future Energy, Fudan University, Shanghai, 200433, China.}
%     \IEEEauthorblockA{\IEEEauthorrefmark{3}The Department of Electronic Engineering, Tsinghua University, Beijing, China.}
% }

\author{
    % \IEEEauthorblockN{
    Xuesong~Wang, Xinyan~Xie, and Runxin~Zhang   
    % \\
    % {wangxuesong@cuhk.edu.cn}\IEEEauthorrefmark{4}
    % }
    % \IEEEauthorblockA{\IEEEauthorrefmark{1}The Chinese University of Hong Kong, Shenzhen, 518172, China.}
    % \IEEEauthorblockA{\IEEEauthorrefmark{2}College of Smart Materials and Future Energy, Fudan University, Shanghai, 200433, China.}
    % \IEEEauthorblockA{\IEEEauthorrefmark{3}The Department of Electronic Engineering, Tsinghua University, Beijing, China.}
    
\thanks{
X. Wang is with The Chinese University of Hong Kong, Shenzhen, Guangdong, China.
X. Xie is with College of Smart Materials and Future Energy, Fudan University, Shanghai, China.
R. Zhang is with the Department of Electronic Engineering, Tsinghua University, Beijing, China.}
\thanks{Corresponding author: Xuesong Wang (wangxuesong@cuhk.edu.cn).
}
% \thanks{ % (rxz@mail.tsinghua.edu.cn)
% }
}

\markboth{}
{Tube-Structured Incremental Semantic HARQ for Generative Video Receivers}
% {Incremental Semantic HARQ for Generative Video Receivers with Tube-Structured Requests}
% 题目换一下，tube-structured往前提

\IEEEaftertitletext{\vspace{-1.2em}}
\maketitle

\begin{abstract}
Generative semantic communication uses receiver-side generative priors to reconstruct visual content from compact semantics, making it attractive for bandwidth-limited multimedia delivery. For video, reliable recovery remains difficult because errors accumulate over time, useful evidence is temporally correlated, and the receiver must make decisions under limited interaction, retransmission, and reconstruction budgets. Existing generative semantic communication studies mainly emphasize representation, compression, or generative reconstruction, while recent error-resilient and semantic-HARQ methods still largely operate on encoder-defined or frame-block retransmission units. This paper studies receiver-driven semantic HARQ for generative video reconstruction under a budget-constrained AoIS-AUC objective and argues that the retransmission primitive is itself an important system design variable. We propose tube-structured package-native requests, in which temporally local packages are the channel-visible HARQ objects and are transmitted, dropped, received, and committed at package granularity. Under a controlled comparison protocol with matched backbone, budgets, and channel model, this primitive yields lower time-weighted recovery cost than competitive block-based baselines in practically relevant moderate-to-harsh regimes, while the gap naturally shrinks in near-clean channels. The gain mainly appears as earlier stabilization of the recovery trajectory, while final-quality endpoints remain broadly comparable, and it persists even against a tube-aware block-ranking baseline.
\end{abstract}


% \vspace{-0.1in}
\section{Introduction}
% \vspace{-0.05in}

% Semantic communication moves wireless systems beyond bit fidelity toward task-oriented and meaning-aware transmission \cite{Qin2023GeneralizedSemCom}. In generative semantic communication, 
% % strong 
% receiver-side priors make it possible to reconstruct visual content from compact semantics and limited side information \cite{Liang2024GenAISemComSurvey, Ren2025GenerativeSemComSurvey}. Recent studies further show that diffusion-based transmission and receiver-driven interaction can support progressive semantic recovery under constrained bandwidth \cite{Guo2024DiffusionDrivenSemCom, Wang2025DiffSemAdaptiveRequests}. For video semantic transmission, however, the challenge is no longer only how to represent semantics compactly. Because video recovery must handle temporal error accumulation and the persistence of missing evidence across frames \cite{DiffuEraser2025,video_inpainting_eval}, the way incremental evidence is requested and delivered over an unreliable link becomes a key part of the system design.
% strong出现太多，可以换几个不同的词

% {\color{red}
Semantic communication shifts the focus of wireless systems from bit-level fidelity toward task-oriented and meaning-aware transmission \cite{yang2022semantic}. Within this paradigm, generative semantic communication utilizes receiver-side priors to reconstruct high-fidelity content from highly compact latent representations \cite{Liang2024GenAISemComSurvey,Ren2025GenerativeSemComSurvey}. While diffusion-based models have successfully enabled progressive recovery for images \cite{Guo2024DiffusionDrivenSemCom,Wang2025DiffSemAdaptiveRequests}, video transmission introduces distinct complexities beyond simple compression. Specifically, the inter-frame persistence of missing evidence and subsequent temporal error accumulation necessitate a shift in focus from representation to interaction \cite{video_inpainting_eval,DiffuEraser2025}. In these scenarios, the mechanism for requesting and delivering the incremental evidence emerges as a critical yet under-explored component of system design.

Advances in generative models have established the feasibility of high-quality video delivery and restoration even from partial observations \cite{Yin2025GenerativeVideoSemCom,Li2025GoalOrientedVideoGenAI,DiffuEraser2025}. 
Building upon these generative backbones, recent research has explored various error-resilient mechanisms, including packetization with generative recovery under loss \cite{Huang2025GenerativeFeatureImputing}, semantic-aware hybrid automatic repeat request (HARQ) design \cite{Hu2024SemHARQ,Li2025SemanticHARQITS}, semantic-aware and QoE-guided resource allocation \cite{Zhang2023IntelligentResourceAllocation,Yan2023QoEBasedResourceAllocation}, and receiver-side adaptive refinement \cite{Wang2025DiffSemAdaptiveRequests}.
However, these methods typically optimize encoder-defined semantic units or network-level control variables while presupposing a fixed channel-visible transport object. Such constraints hinder generative video recovery, as the transport primitive governs the temporal distribution of evidence arrival. Isolated frame-block transport fails to exploit temporal coherence, leaving the receiver semantically starved and stalled in an incorrect state until enough spatiotemporal support has trickled in to resolve the reconstruction. Therefore, this work identifies the retransmission primitive as a critical design variable for optimizing generative video, referred to as incremental semantic HARQ.

% Age-aware metrics provide a suitable perspective for this issue, because the main question is not only the final reconstruction endpoint, but also how quickly the receiver can leave an incorrect state. Age of incorrect information (AoII) and its semantic extensions have been widely used to characterize persistent incorrectness and delayed recovery in communication systems. \cite{Maatouk2023AoIIEnabler, Han2026AoISRemoteMonitoring, bountrogiannis2025aoii_harq
% % ,Chen2025AgeOfSemantics
% } In this work, we study receiver-driven semantic HARQ for generative video reconstruction under a budget-normalized area-under-curve objective based on age of incorrect semantics (AoIS-AUC). Our system uses diffusion inpainting as the fixed generative receiver backbone \cite{DiffuEraser2025}. Specifically, we propose a tube-structured package-native design that treats temporally local packages as channel-visible HARQ objects, so that evidence is transmitted, dropped, received, and committed at package granularity rather than at isolated frame-block granularity.

To quantify the duration of such incorrect states, we adopt a budget-constrained Area-Under-the-Curve objective based on the Age of Incorrect Semantics (AoIS-AUC) to characterize delayed recovery in generative video reconstruction. Similar to the Age of Incorrect Information (AoII), this metric prioritizes the speed of transition from incorrect to reliable states over final reconstruction fidelity alone \cite{Maatouk2023AoIIEnabler,Han2026AoISRemoteMonitoring,bountrogiannis2025aoii_harq}. Furthermore, leveraging the intrinsic spatiotemporal correlations within video sequences, this paper proposes a tube-structured package as the channel-visible HARQ object. 
Then, we develop a receiver-side greedy request policy to minimize AoIS-AUC and evaluate the proposed framework across varying bandwidth and computational budgets to demonstrate its effectiveness in practical communication scenarios. The main contributions are summarized as follows:

\begin{figure*}[!t]
    \centering
    \includegraphics[width=\linewidth]{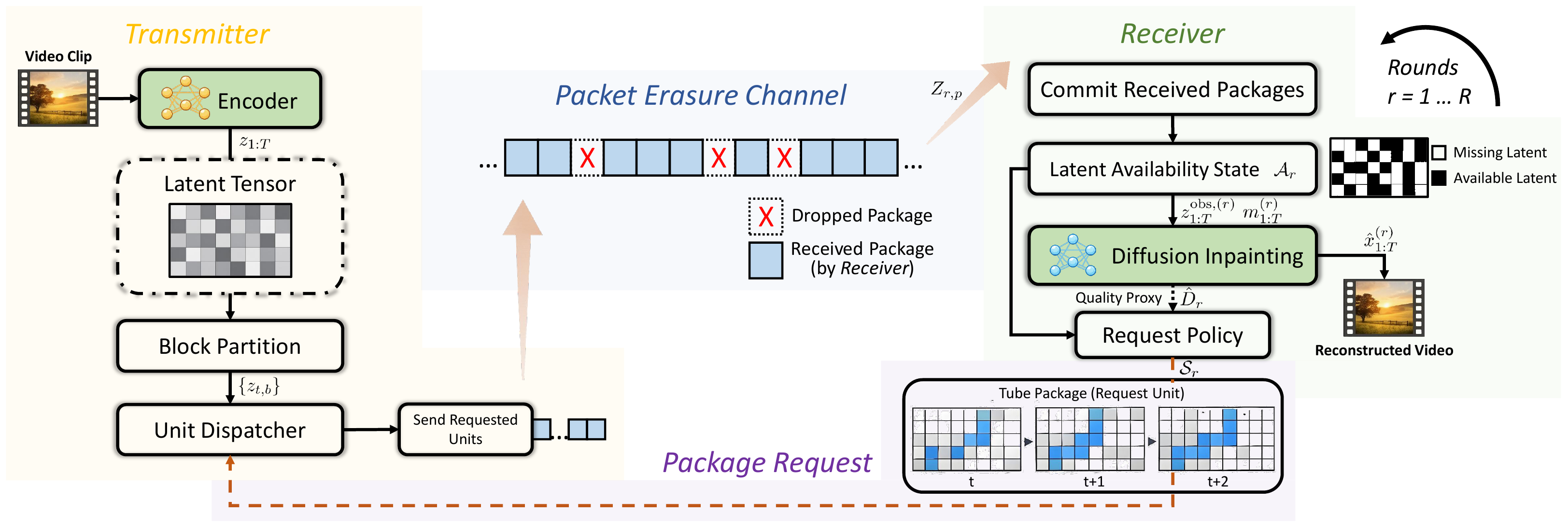}
    \caption{System overview of receiver-driven semantic HARQ for generative video reconstruction. The receiver requests tube-structured packages, the forward link transports them as atomic HARQ objects, and successful deliveries are committed before diffusion inpainting and the next feedback decision.}
    \label{fig:system_primitive}
\end{figure*} 

\vspace{-0.03in}
\begin{itemize}[leftmargin= 0.4cm]
    % \item Unlike traditional schemes that rely on fixed frame-block units, we identify the retransmission primitive itself as a critical system design variable in generative video semantic communication. By projecting object regions onto the latent space, we aggregate spatiotemporal latent blocks into coherent tubes and propose the tube-structured packages as the channel-visible HARQ objects.
    % \item To minimize the AoIS-AUC, we propose a greedy request strategy that prioritizes candidate packages according to their uncovered ratios, mean support areas, and temporal spans. Coupled with a diffusion-based generative model, this approach facilitates a more rapid stabilization of the video recovery trajectory.
    % \item Extensive simulations validate the effectiveness of our proposed scheme, achieving up to a 6-round recovery lead in moderate-to-harsh regimes. Moreover, we offer system-level design guidelines across diverse channel conditions and motion intensities to facilitate practical application.
    \item Unlike traditional schemes relying on fixed frame-block units, we identify the retransmission primitive itself as a critical system design variable in generative video semantic communication. By projecting object regions onto the latent space, we aggregate spatiotemporal latent blocks into coherent tubes and propose tube-structured packages as the channel-visible HARQ objects.
    \item To minimize the AoIS-AUC, we propose a greedy request strategy that prioritizes candidate packages by their uncovered ratios, mean support areas, and temporal spans. Coupled with a diffusion-based generative model, this approach facilitates more rapid stabilization of the video recovery trajectory.
    \item Extensive simulations validate the effectiveness of the proposed scheme, achieving up to a 6-round recovery lead in moderate-to-harsh regimes. Moreover, we offer system-level design guidelines across diverse channel conditions and motion intensities for practical application.
    \end{itemize}
% }

% \begin{itemize}
%     \item We identify retransmission primitive selection as a distinct design issue in receiver-driven generative video HARQ, and show that per-frame block requests become a weak control lever under tight budgets.
%     \item We propose a tube-structured package-native HARQ design in which temporally local packages are the channel-visible HARQ objects, so that the forward link operates on packages rather than isolated frame blocks.
%     \item Under matched backbone, budgets, and channel settings, we show that this primitive improves the time-weighted recovery trajectory in moderate-to-harsh regimes, while final reconstruction quality remains broadly comparable and the gain naturally shrinks in near-clean regimes.
%     % % another version and need further changes
%     % \item We identify the retransmission primitive as a distinct design issue in receiver-driven generative video HARQ, and propose a tube-structured package-native design in which temporally local packages are the channel-visible HARQ objects.
%     % \item Under matched backbone, budgets, and channel settings, we show that this primitive improves the time-weighted recovery trajectory in moderate-to-harsh regimes where interaction remains consequential, while the gain naturally shrinks in near-clean channels.
%     % \item Experiments indicates that package-native transport remains effective in both low and high-motion settings with different gain magnitude, and its benefit appears primarily in faster recovery rather than in endpoint-quality improvement.
% \end{itemize}

\section{System Design and Tube-Structured Package Transport}
%换个title强调设计。以及，tube-struction可以多强调时空属性
\label{sec:system_model}
% x_t以及z_t的改动范围太大，暂时保持原样

We consider receiver-driven semantic HARQ for generative video reconstruction over an unreliable wireless link, as illustrated in Fig.~\ref{fig:system_primitive}. After an initial semantic payload is transmitted, the receiver enters interactive HARQ rounds, tracks latent availability, optionally performs diffusion inpainting, evaluates a quality proxy, and feeds back the next package request. A transmitter-side dispatcher then sends the requested units through the erasure channel. The key design choice is the retransmission primitive that uses temporally local tube-structured packages as the channel-visible HARQ objects, instead of individual frame-blocks.

\subsection{System Flow and Package-Native HARQ}
\label{subsec:system_flow}

% x_{1:T} 加粗表示更好；z_t也是
Let $x_{1:T}=\{x_t\}_{t=1}^T$ denote a video clip, where each frame $x_t\in\mathbb{R}^{C\times H\times W}$. A pretrained encoder $E(\cdot)$ maps each frame to a latent tensor $z_t = E(x_t) \in \mathbb{R}^{C_L \times H_L \times W_L}$.
% \begin{equation}
% \mathbf{z}_t = E(x_t) \in \mathbb{R}^{C_L \times H_L \times W_L}, \\qquad t=1,\dots,T.
% \end{equation}
Each latent tensor $z_t$ is partitioned into $N$ spatial blocks with resulting blocks denoted by $\{z_{t,b}\}_{b=1}^{N}$. The full spatiotemporal latent block set is
\begin{equation}
\mathcal{U} \triangleq \{(t,b): t\in\{1,\dots,T\},\; b\in\{1,\dots,N\}\}.
% 这里的b换个字母，比如n
\end{equation}

For each clip, an offline \textit{tube-extraction stage} constructs a fixed package catalog $\mathcal{C}=\{p\}_{p=1}^{P}$. % tube-extraction stage斜体
% \mathcal{C}和前面的x_{1:T}；z_t统一一下，比如都是花体
Each package $p$ is associated with a member set $\mathcal{P}_p\subseteq\mathcal{U}$, an owner tube index $o(p)$, a temporal span $\ell_p$ equal to the number of frames covered by $\mathcal{P}_p$, and a package size in budget units $c_p\triangleq |\mathcal{P}_p|$.
Tube packages are constructed once per clip from frame-wise object regions projected onto the latent grid, that is, each object first induces a spatiotemporal latent-block tube, which is then split into temporally local packages under fixed size constraints; overlap is resolved before finalizing the catalog, and background packages cover the remaining latent blocks. 
The final catalog is non-overlapping and complete:
\begin{equation}
\mathcal{U} = \bigcup_{p\in\mathcal{C}} \mathcal{P}_p,
\qquad
\mathcal{P}_p \cap \mathcal{P}_{p'} = \emptyset,\ \forall p\neq p'.
\label{eq:catalog_partition_main}
\end{equation}
Hence, the receiver ultimately reconstructs from latent blocks, but the forward link operates on packages rather than individual blocks.

Let $\mathcal{A}^{\mathrm{init}}\subseteq\mathcal{U}$ denote the latent blocks made available by the initial semantic payload. In each interactive HARQ round $r=1,\dots,R$, the receiver feeds back a package request 
% $\mathcal{S}_r \subseteq \mathcal{C},\sum_{p\in\mathcal{S}_r} c_p \le K$,
\begin{equation}
\mathcal{S}_r \subseteq \mathcal{C},
\qquad
\sum_{p\in\mathcal{S}_r} c_p \le K, % c_p可以换个字母
\label{eq:req_budget_main}
\end{equation}
where $K$ is the per-round request budget. The feedback is assumed reliable and contains only package identifiers.

The forward link is modeled as a packet-erasure channel over the requested channel-visible units. In the main experiments, we use a Gilbert-Elliott (GE) burst-erasure channel. 
Let $e_i\in\{0,1\}$ denote the erasure indicator of the $i$-th transmitted unit in chronological order, where $e_i=1$ indicates erasure. The state evolution follows
\begin{equation}
\Pr(e_i=1\!\mid\!e_{i-1}\!=\!0)\!=\!p_{01},~
\Pr(e_i=0\!\mid\!e_{i-1}\!=\!1)\!=\!p_{10}.
\label{eq:ge_transition_main}
\end{equation}
% To match an average packet erasure rate (PER) and an average burst length $L$, we set
% \begin{equation}
% p_{10}=\frac{1}{L},
% \qquad
% p_{01}=\frac{\mathrm{PER}}{1-\mathrm{PER}}\,p_{10}.
% \label{eq:ge_match_per_main}
% \end{equation}
To match a target packet erasure rate (PER) and burst length $L$, we nominally set
\begin{equation}
p_{10}=\frac{1}{L},
\qquad
p_{01}=\frac{\mathrm{PER}}{1-\mathrm{PER}}\,p_{10},
\label{eq:ge_match_per_main}
\end{equation}
and reduce $p_{10}$ accordingly if $p_{01}\ge 1$ so that the target PER is preserved while the effective burst length becomes larger.

For each requested package $p\in\mathcal{S}_r$, let $Z_{r,p}\in\{0,1\}$ indicate successful delivery. Package delivery is atomic: if $Z_{r,p}=1$, all members of $\mathcal{P}_p$ are committed; otherwise none are. Define $\mathcal{A}_0 \triangleq \mathcal{A}^{\mathrm{init}}$; throughout this paper, the superscript $\mathrm{init}$ and the index $0$ are used equivalently unless otherwise stated. The latent availability state is then updated by
\begin{equation}
\mathcal{A}_r
=
\mathcal{A}_{r-1}
\cup\Biggl(
\bigcup_{p\in\mathcal{S}_r:\, Z_{r,p}=1}
\mathcal{P}_p\Biggr),
\qquad r=1,\dots,R.
\label{eq:availability_update_main}
\end{equation}
This is the essential structural difference from block-native transport, where HARQ operates on individual frame-blocks rather than packages.

Given $\mathcal{A}_r$, define the missing set as % \mathds{1}定义一下
\begin{equation}
\mathcal{M}_r \triangleq \mathcal{U}\setminus \mathcal{A}_r,
\qquad
m_{t,b}^{(r)} \triangleq \mathds{1}\!\left[(t,b)\in \mathcal{M}_r\right],
\end{equation}
and masked latent observation as
\begin{equation}
z_{t,b}^{\mathrm{obs},(r)}
\triangleq
z_{t,b}\,\mathds{1}\!\left[(t,b)\in \mathcal{A}_r\right].
\end{equation}
Here $\mathds{1}[\cdot]$ denotes the indicator function. Collecting these entries over all $(t,b)$, we write $z_{1:T}^{\mathrm{obs},(r)}$ and $m_{1:T}^{(r)}$ for the masked latent observation and the corresponding mask. The receiver reconstructs the clip by diffusion inpainting
\begin{equation}
\hat{x}_{1:T}^{(r)}
=
G\!\left(z_{1:T}^{\mathrm{obs},(r)},\, m_{1:T}^{(r)} \right),
\label{eq:reconstruction_main}
\end{equation}
where $G(\cdot)$ is the pretrained generative receiver. The initial-stage reconstruction is denoted by $\hat{x}_{1:T}^{\mathrm{init}}$.

Let $u^{\mathrm{init}}\in\{0,1\}$ and $u_r\in\{0,1\}$ denote the reconstruction decisions in the initial stage and in round $r$, respectively. Under a total compute budget (CB) $b_c$, we have
%后面的CB就是这个，不要重复定义
\begin{equation}
u^{\mathrm{init}} + \sum_{r=1}^{R} u_r \le b_c.
\label{eq:compute_budget_main}
\end{equation}
The elapsed time is modeled by separating the one-shot initial cost from the round-wise interaction cost. Here, $t^{\mathrm{init}}$ is the elapsed time of the initial stage, $\Delta_r$ is the time increment in round $r$, and $t_r$ is the cumulative elapsed time up to round $r$; $c_{\mathrm{init}}$, $c_{\mathrm{RTT}}$, $c_{\mathrm{pkt}}$, and $c_{\mathrm{inp}}$ denote the initial transmission, feedback, per-packet transmission, and reconstruction costs, respectively, i.e.,
% 几个cost都用一个字母c显得奇怪
\begin{equation}
t^{\mathrm{init}}
\triangleq
c_{\mathrm{init}} + c_{\mathrm{inp}}\,u^{\mathrm{init}},
\label{eq:initial_time_main}
\end{equation}
\begin{equation}
\Delta_r
=c_{\mathrm{RTT}}+c_{\mathrm{pkt}} \sum_{p\in\mathcal{S}_r} c_p
+c_{\mathrm{inp}} u_r,~~r=1,\dots,R,
\label{eq:time_increment_main}
\end{equation}
and
\begin{equation}
t_r \triangleq t^{\mathrm{init}} + \sum_{i=1}^{r} \Delta_i.
\label{eq:time_update_main}
\end{equation}

\subsection{Tube-Package AoIS-AUC Objective}
\label{subsec:tube_pipeline_aoii}

Our goal is early distortion reduction without sacrificing final reconstruction quality.
Let
\begin{equation}
D^{\mathrm{init}}
=\phi\!\left(x_{1:T},\hat{x}_{1:T}^{\mathrm{init}}\right), ~~
D_r
=\phi\!\left(x_{1:T},\hat{x}_{1:T}^{(r)}\right),
% r=1,\dots,R.
\label{eq:distortion_main}
\end{equation}
where $\phi(\cdot)$ is a fixed distortion metric. 
Since the receiver does not observe the ground-truth clip during interaction, $D^{\mathrm{init}}$ and $D_r$ are used for offline evaluation, while online control relies on a shared receiver-side proxy $\hat{D}^{\mathrm{init}}$ and $\hat{D}_r$ derived from the current missing-state structure and burst-loss state, with lightweight calibration when reconstruction feedback becomes available.

Let $t_0\triangleq t^{\mathrm{init}}$. 
% We define the distortion trajectory as piecewise constant over time $D(t)=D^{\mathrm{init}}, t\in(0,t_0]$ and $D(t)=D_r, t\in(t_{r-1},t_r]$.
We define the distortion trajectory as piecewise constant over continuous time $\tau$, with $D(\tau)=D^{\mathrm{init}}$ for $\tau\in(0,t_0]$ and $D(\tau)=D_r$ for $\tau\in(t_{r-1},t_r]$.
The resulting age-weighted recovery objective is
\begin{equation}
\hspace{-0.2cm} J_{\mathrm{AoIS}}
\triangleq
% \int_0^{t_R} t\,D(t)\,dt
\int_0^{t_R} \tau\,D(\tau)\,d\tau
=
\frac{t_0^2}{2}\,D^{\mathrm{init}}
+
\sum_{r=1}^{R}
\frac{t_r^2-t_{r-1}^2}{2}\,D_r.
\label{eq:aoiis_auc_main_sys}
\end{equation}
Therefore, when two methods reach similar final quality, the one that reduces distortion earlier attains a smaller objective value. This objective defines what the receiver-side policy should optimize under the request and compute budgets introduced above.

\addtolength{\topmargin}{0.05in}
\begin{algorithm}[t]
\caption{Receiver-side greedy package request policy}
\label{alg:request_policy_main}
\begin{algorithmic}[1]
\Require $\mathcal{A}^{\mathrm{init}}$, $\mathcal{C}$, $K$, $b_c$, $u^{\mathrm{init}}$, $\hat{x}_{1:T}^{\mathrm{init}}$, $\hat{D}^{\mathrm{init}}$
\For{$r=1,\dots,R$}
    \State Compute $\rho_{p,r}$ for all $p\in\mathcal{C}$ and form $\mathcal{V}_r^{\mathrm{pkg}}$ by (\ref{eq:active_package_candidates_main}) % Eq.除了句首都不加
    \State Score each $p\in\mathcal{V}_r^{\mathrm{pkg}}$ by (\ref{eq:package_score_main})
    \State Rank $\mathcal{V}_r^{\mathrm{pkg}}$ by descending $s_{p,r}$ and greedily form $\mathcal{S}_r$ under (\ref{eq:req_budget_main})
    \State Transmit all $p\in\mathcal{S}_r$, observe $Z_{r,p}$, and update $\mathcal{A}_r$ by (\ref{eq:availability_update_main})
    \State Set $u_r$ by (\ref{eq:compute_trigger_main})    
    \If{$u_r=1$}
        \State Set $\hat{x}_{1:T}^{(r)}\!\leftarrow \!G(z_{1:T}^{\mathrm{obs},(r)},\!m_{1:T}^{(r)})$ and update $\hat{D}_r$
    \Else
        \State Set $\hat{x}_{1:T}^{(r)} \leftarrow \hat{x}_{1:T}^{(r-1)}$ and $\hat{D}_r \leftarrow \hat{D}_{r-1}$
    \EndIf
\EndFor
\end{algorithmic}
\end{algorithm}

\subsection{Request Policy}
\label{subsec:request_policy_main}

The proposed request policy is a receiver-side greedy policy operating on the fixed package catalog $\mathcal{C}$. At round $r$, the receiver first evaluates the uncovered fraction of each package: % how much不准确
\begin{equation}
\rho_{p,r}
\triangleq
\frac{|\mathcal{P}_p \cap (\mathcal{U}\setminus\mathcal{A}_{r-1})|}{|\mathcal{P}_p|},
\qquad p\in\mathcal{C}.
\label{eq:uncovered_ratio_main}
\end{equation}
Only packages with $\rho_{p,r}\!\!>\!\!0$ are considered as candidates. Let
\begin{equation}
\mathcal{V}_r^{\mathrm{pkg}}
\triangleq
\{p\in\mathcal{C}:\rho_{p,r}>0\}.
\label{eq:active_package_candidates_main}
\end{equation}

Each candidate package $p\in\mathcal{V}_r^{\mathrm{pkg}}$ is then scored by
\begin{equation}
s_{p,r}
=
w_1 \rho_{p,r}
+
w_2 \bar{A}_{o(p)}
+
w_3 \ell_p,
\label{eq:package_score_main}
\end{equation}
% where $o(p)$ is the owner tube of package $p$, $\bar{A}_{o(p)}$ is the mean latent support area of that tube, and $\ell_p$ is the temporal span defined in Sec. \ref{subsec:system_flow}.
where $o(p)$ denotes the tube that generated package $p$, $\bar{A}_{o(p)}$ is the mean number of latent blocks covered per frame by that tube, and $\ell_p$ is the temporal span defined in Sec. \ref{subsec:system_flow}.
The score weights $(w_1,w_2,w_3)$ are fixed throughout the experiments. The first term favors immediate uncovered utility, while the latter two provide a lightweight bias toward larger and temporally longer tube-supported evidence.

The request set $\mathcal{S}_r$ is formed by ranking $\mathcal{V}_r^{\mathrm{pkg}}$ in descending score order and greedily selecting candidates under the request-budget constraint in (\ref{eq:req_budget_main}). The reconstruction trigger is shared by the three main online methods and depends on the current proxy distortion and the remaining compute budget. Specifically,
\begin{equation}
u_r
=
\mathds{1}\!\left[\hat{D}_{r-1}\ge \tau_{\mathrm{trig}}\right]
\cdot
\mathds{1}\!\left[b_c-u^{\mathrm{init}}-\sum_{i=1}^{r-1}u_i > 0\right],
\label{eq:compute_trigger_main}
\end{equation}
where $\tau_{\mathrm{trig}}$ is a fixed trigger threshold. Algorithm~\ref{alg:request_policy_main} summarizes the round-wise request, commitment, and reconstruction procedure.

\section{Experiments and Results}
\label{sec:experiments_results}

\subsection{Experimental Setup, Baselines, and Metrics} % table-1没有cite
\label{subsec:exp_setup_baselines}

Experiments are conducted on the 480p trainval split of DAVIS 2017 dataset, using DiffuEraser as the pretrained generative receiver. 
The main experimental settings are listed in Table \ref{tab:main_exp_params}. The tight computing budget $b_c\in\{2,3\}$ reflects a low-update regime in which only a few reconstructions are allowed over the full HARQ interaction, and each reconstruction requires a full diffusion inference pass, making reconstruction the dominant compute cost \cite{hu2025easyomnimatte}.
The main regime plots use $K\in\{8,16\}$, which keeps interaction active while maintaining meaningful request selection under limited budgets.
The evaluated methods are as follows.
\begin{itemize}[leftmargin = 0.4cm]
  \item \textbf{Tube-Package Requests (Ours).} The receiver requests temporally local packages as atomic HARQ objects.
  \item \textbf{Greedy Block Requests (Baseline).} A block-native baseline that requests currently missing frame-blocks.
  \item \textbf{Tube-Weighted Block Requests (Baseline).} A stronger block-native baseline that already uses tube information for ranking while still transporting individual blocks, thereby controlling for tube-aware ordering.
  % \item \textbf{Hysteresis Trigger (Ref.).} A trigger-side reference included only for contextual comparison.
  % \item \textbf{Offline Oracle (Ref.).} An offline upper-bound reference under the same budgets. 
  \item \textbf{Hysteresis Trigger (Ref.).} A block-native reference that keeps the same missing-block request primitive as Greedy Block Requests, but replaces the reconstruction trigger with a high/low-threshold hysteresis rule.
  \item \textbf{Offline Planning (Ref.).} A noncausal reference that keeps the same budget constraints but plans reconstruction timing offline using a simplified surrogate objective.
  % upper-bound为什么是最差的？而且并不是最值。而且oracle目前没有实心和空心
\end{itemize}

\vspace{0.2em}
\begin{table}[!t]
\centering
\caption{Main experimental parameters.}
\label{tab:main_exp_params}
\renewcommand{\arraystretch}{1.15}
\setlength{\tabcolsep}{10pt}
\footnotesize
\begin{tabular}{l|c|l}
\hline
\makecell[c]{\textbf{Item}} & \textbf{Symbol} & \makecell[c]{\textbf{Value}} \\
\hline
HARQ horizon & $R$ & $6$ \\
Compute budget  & $b_c$ & $\{2,3\}$ \\
Request budget & $K$ & $\{8,16\}$ \\
Mean burst length & $L$ & $4$ \\
Trigger threshold & $\tau_{\mathrm{trig}}$ & $0.35$ \\
RTT cost & $c_{\mathrm{RTT}}$ & $0.01$ \\
Packet cost & $c_{\mathrm{pkt}}$ & \makecell[l]{$1.024 \times 10 ^{- 4}$ \\(1024-bit packet at 10 Mbps)} \\
Reconstruction cost\!\! & $c_{\mathrm{inp}}$ & $3.0$ \\
Package span & $\ell_p$ & at most $3$ frames \\
Package size & $|\mathcal{P}_p|$ & $4$-$24$ latent blocks \\
Score weights & \!\!\!$(\!w_1\!,\!w_2\!,\!w_3\!)$\!\!\! & $(1.0,0.5,0.25)$ \\
\hline
\end{tabular}
\end{table}

% The first three methods share the same initial semantic payload, generative receiver, distortion evaluator, interaction horizon, budgets, GE channel setting, compute-trigger rule, and time-cost accounting. Therefore, the intended difference remains the request/transport primitive rather than hidden changes in backbone capacity, trigger behavior, or resource accounting.

% The first three methods form the main online comparison, while hysteresis trigger and offline planning are included only as reference methods for context. For the two block-based baselines, the channel-visible unit is an individual latent block, whereas for the proposed method it is an atomic package. Unless otherwise stated, the primary metric is AoIS-AUC, where lower is better. We also report recovery delay

The first three methods form the main online comparison and share the same initial semantic payload, generative receiver, distortion evaluator, interaction horizon, budgets, GE channel setting, compute-trigger rule, and time-cost accounting. Thus, the intended difference is only the request/transport primitive, that is, the two block-based baselines operate on individual latent blocks, whereas the proposed method operates on atomic packages. Hysteresis trigger and offline planning are included only as contextual reference methods. Unless otherwise stated, the primary metric is AoIS-AUC, where lower is better. We also report recovery delay
% \cite{Maatouk2023AoIIEnabler
% ,Maatouk2020AoII
% }
\begin{equation}
t_{\alpha}
\triangleq
\min\{t_r: D_r \le \alpha D^{\mathrm{init}}\},
~~
t_{\alpha}=t_R \ \text{otherwise},
\label{eq:recovery_delay_main}
\end{equation}
for a fixed threshold fraction $\alpha\in(0,1)$, where smaller $t_\alpha$ means faster stabilization, and the clip-motion score
\begin{equation}
\mu(x_{1:T})
\triangleq
\frac{1}{T-1}\sum_{t=2}^{T}\frac{1}{CHW}\|x_t-x_{t-1}\|_1,
\label{eq:motion_score_main}
\end{equation}
whose median is used to split the dataset into low-motion and high-motion subsets.

\begin{figure}[!t]
  \centering
  \includegraphics[width=\linewidth,trim=0 90bp 0 0,clip]{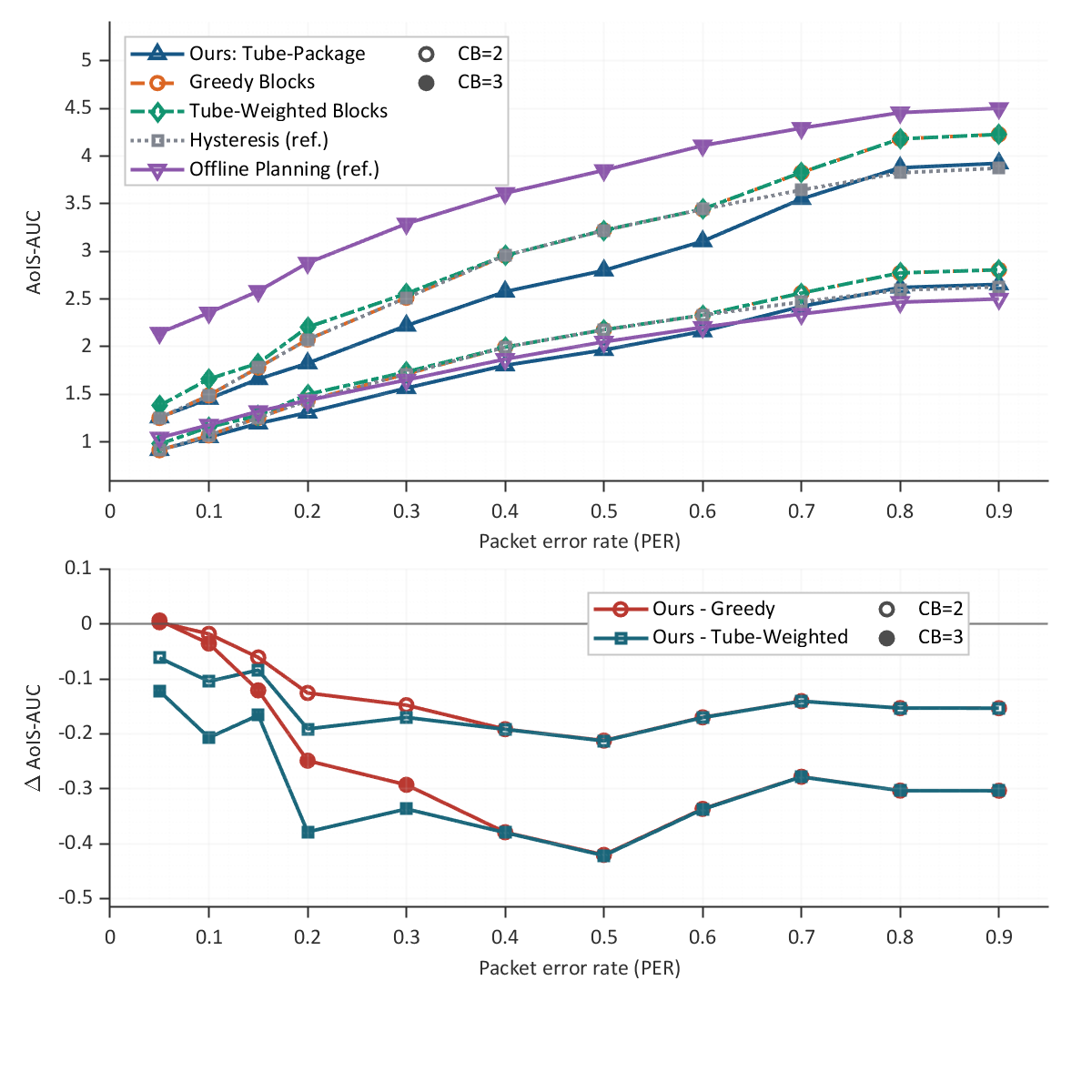}
  \caption{PER sweep under the GE packet-erasure channel at $K=8$, with absolute AoIS-AUC curves (top) and paired gaps to the two block-based baselines (bottom). Hollow and filled markers denote $b_c = 2$ and $b_c = 3$, respectively.}
  \label{fig:per_sweep_k8}
\end{figure}

\begin{figure}[!t]
  \centering
  \includegraphics[width=\linewidth,trim=0 75bp 0 0,clip]{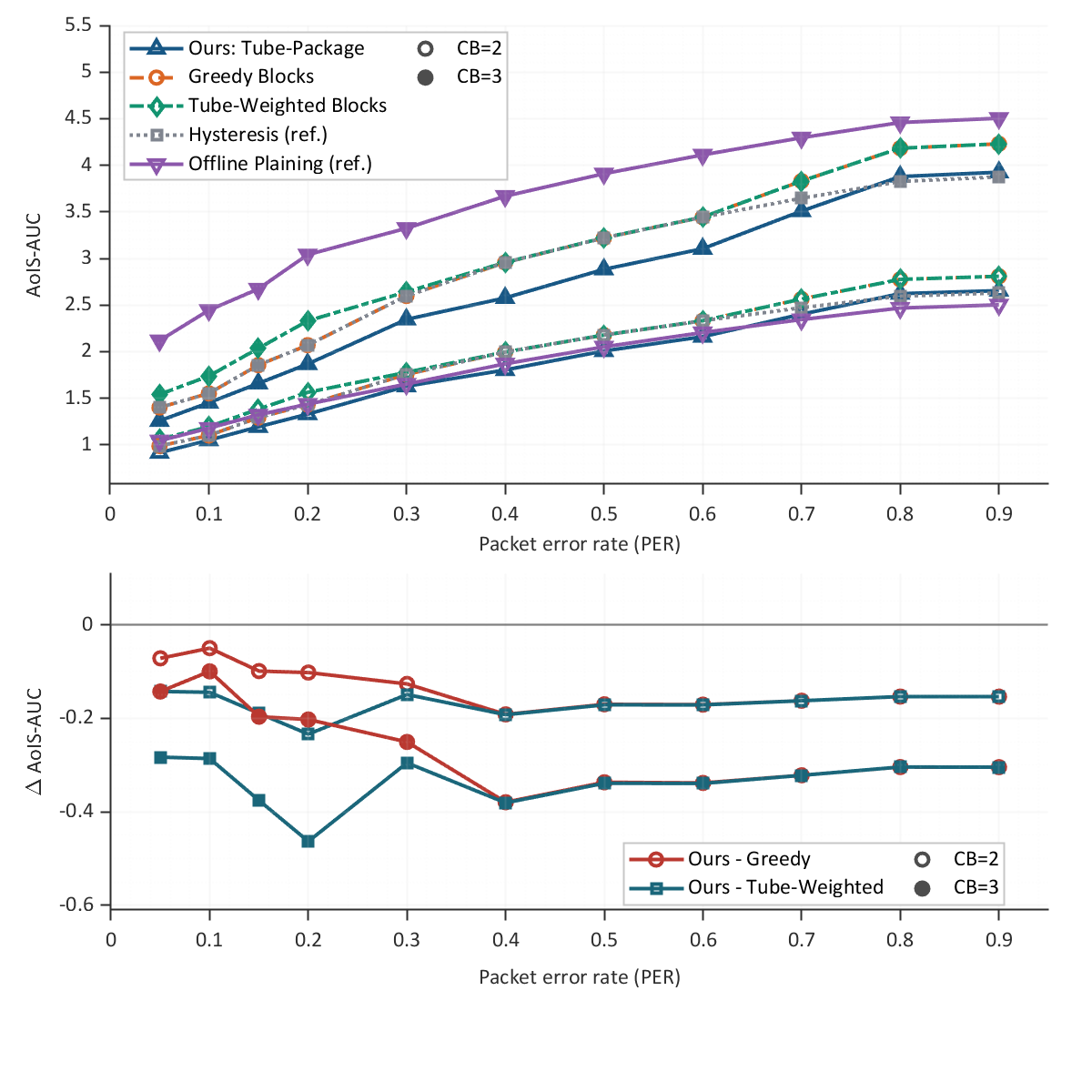}
  \caption{PER sweep of AoIS-AUC at $K=16$.}
  \label{fig:per_sweep_k16}
\end{figure}

\subsection{Results and Discussion}
\label{subsec:results_discussion}

Fig.~\ref{fig:per_sweep_k8} and Fig.~\ref{fig:per_sweep_k16} are the main regime plots. The top panels show absolute AoIS-AUC, and the bottom panels report
$\Delta J_{\mathrm{AoIS}} \triangleq J_{\mathrm{AoIS}}^{\mathrm{ours}} - J_{\mathrm{AoIS}}^{\mathrm{baseline}}$
against the two block-based baselines, where more negative values indicate larger improvement. At $K=8$, the gain is already clear over a broad moderate-PER range. In near-clean channels, it is small against greedy block requests but already negative against tube-weighted block requests. At $K=16$, the negative gap becomes more uniform across the full PER sweep for both $b_c=2$ and $b_c=3$.

Two observations are most important. First, across both $b_c=2$ and $b_c=3$, the proposed primitive stays below tube-weighted block requests over a broad moderate-PER range for both $K=8$ and $K=16$. Since tube-weighted block requests already use tube information for ranking, this remaining gap suggests that the benefit does not reduce to tube-aware ordering alone. Second, the gain over greedy block requests is already broad at $K=8$ and becomes more uniform at $K=16$, so the effect is not tied to a single request-budget choice within the low-budget regime considered here. The same qualitative pattern appears at both compute budgets, while the larger budget often yields a larger absolute gain.

\begin{figure}[!t]
  \centering
  \includegraphics[width=\linewidth,trim=0 15bp 0 0,clip]{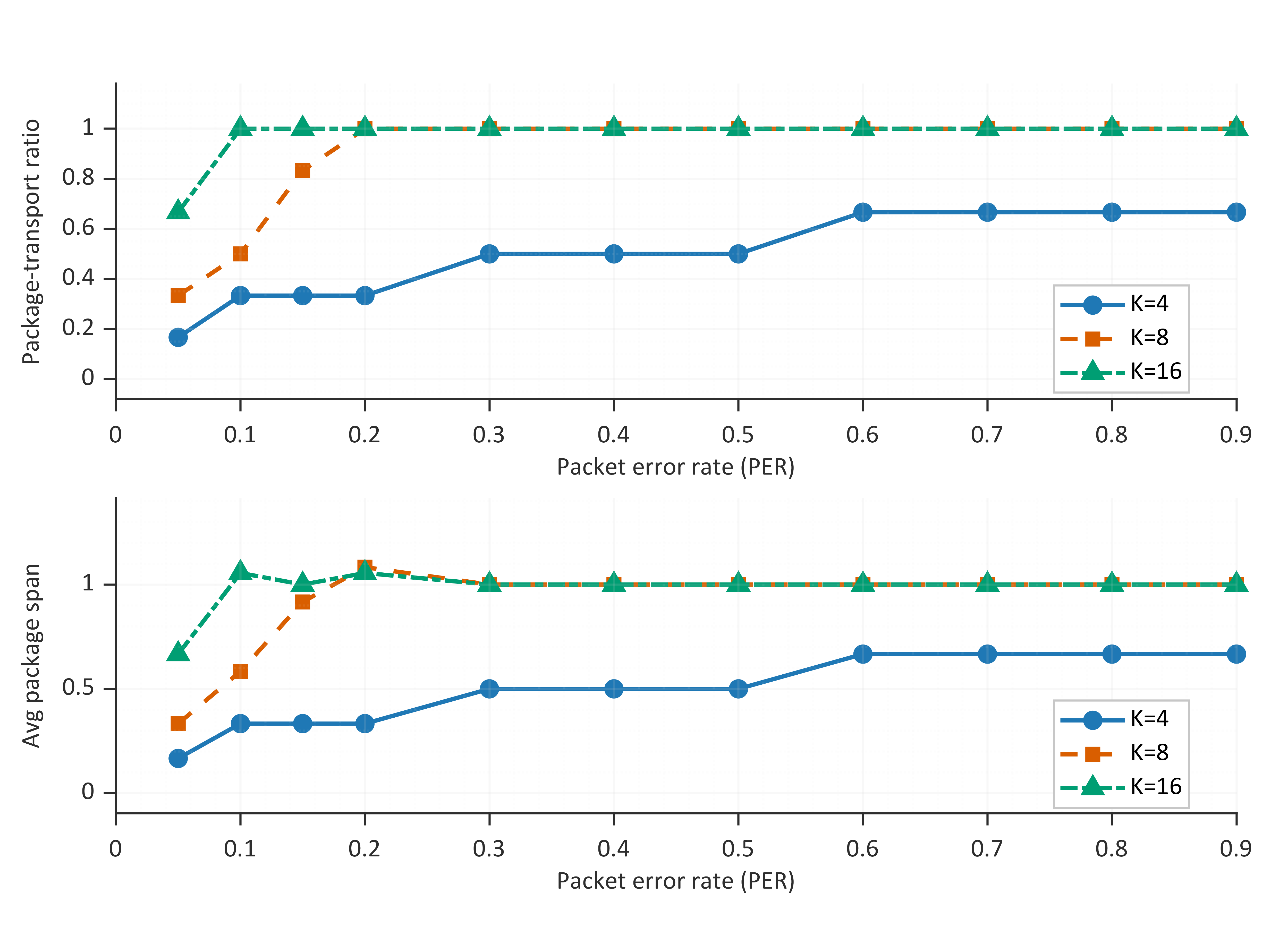}
  \caption{Protocol audit under PER sweeps at $b_c=2$: package-transport ratio and average package span for $K\in\{4,8,16\}$.}
  \label{fig:audit_behavior}
\end{figure}

Fig.~\ref{fig:audit_behavior} helps explain the regime dependence of the gain. For $K=8$, the package-transport ratio rises from about $0.33$ at $\mathrm{PER}=0.05$ to about $0.83$ at $\mathrm{PER}=0.15$, and only saturates near $1$ from $\mathrm{PER}=0.20$ onward; the average package span shows the same transition toward about $1$. Thus, in near-clean channels, the delivered incremental evidence is still too limited for the package primitive to create a clear gap. For $K=16$, by contrast, the transport ratio is already about $0.67$ at $\mathrm{PER}=0.05$ and saturates near $1$ from $\mathrm{PER}=0.10$ onward, with the average package span already close to $1$ in the same range. This earlier saturation is consistent with the more uniform negative $\Delta J_{\mathrm{AoIS}}$ in Fig.~\ref{fig:per_sweep_k16}, whereas the gain at $K=8$ emerges later in Fig.~\ref{fig:per_sweep_k8}.

\begin{figure}[!t]
  \centering
  \includegraphics[width=\linewidth,trim=0 25bp 0 0,clip]{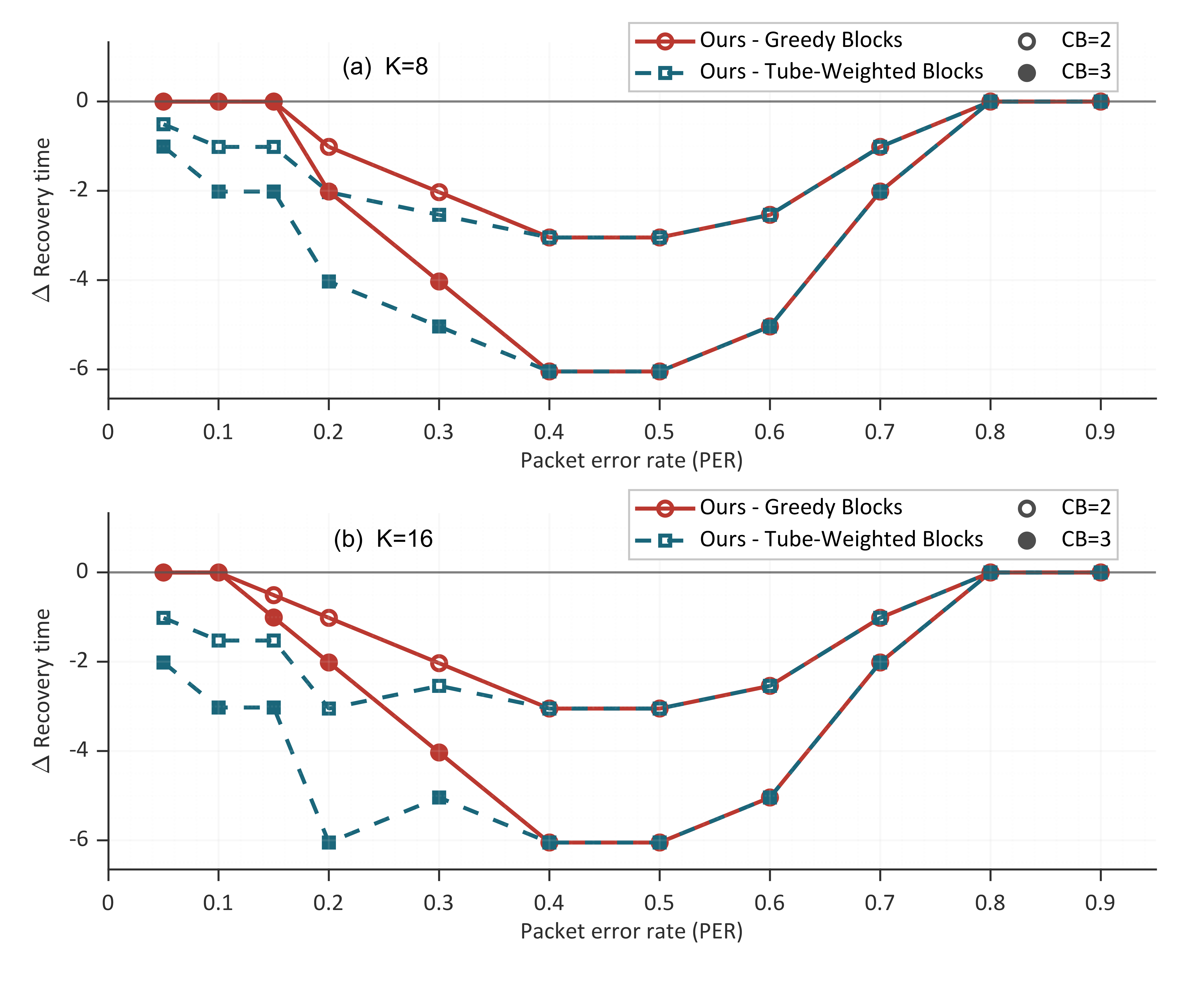}
  \caption{Recovery-time gap under PER sweeps for $K\in\{8,16\}$ and $b_c\in\{2,3\}$. More negative values indicate earlier threshold crossing.}
  \label{fig:recovery_delay}
\end{figure}

Fig.~\ref{fig:recovery_delay} converts the AUC gain into a time-to-reliability view. Here the plotted quantity is the recovery-time gap relative to each baseline, so more negative values indicate earlier threshold crossing for the proposed method. In the moderate-PER regime, this gap is often several rounds and can approach $6$ rounds in the stronger cases, especially at $b_c=3$. Relative to greedy block requests, it is small in very clean channels and strongly negative in the moderate regime; relative to tube-weighted block requests, it is already negative at low PER and reaches its largest magnitude in the moderate regime before collapsing toward zero only in very harsh channels. This links the AoIS-AUC improvement to earlier stabilization rather than to a purely aggregate effect.

% Fig.~\ref{fig:motion_stratification} shows that the improvement is not confined to a special subset of clips. The benefit appears in both motion strata, although its magnitude varies with motion level and baseline choice. Against greedy block requests, the high-motion stratum often shows an earlier onset or slightly larger gains in part of the moderate-PER range, which is consistent with the temporal-coherence argument. Against tube-weighted block requests, the low/high-motion gap is smaller and can reverse at some PER points, suggesting that part of the motion signal is already captured by tube-aware ordering. The appropriate conclusion is therefore not motion exclusivity, but motion-modulated gain with the overall primitive advantage unchanged.

\begin{figure}[!t]
  \centering
  \includegraphics[width=\linewidth,trim=0 65bp 0 0,clip]{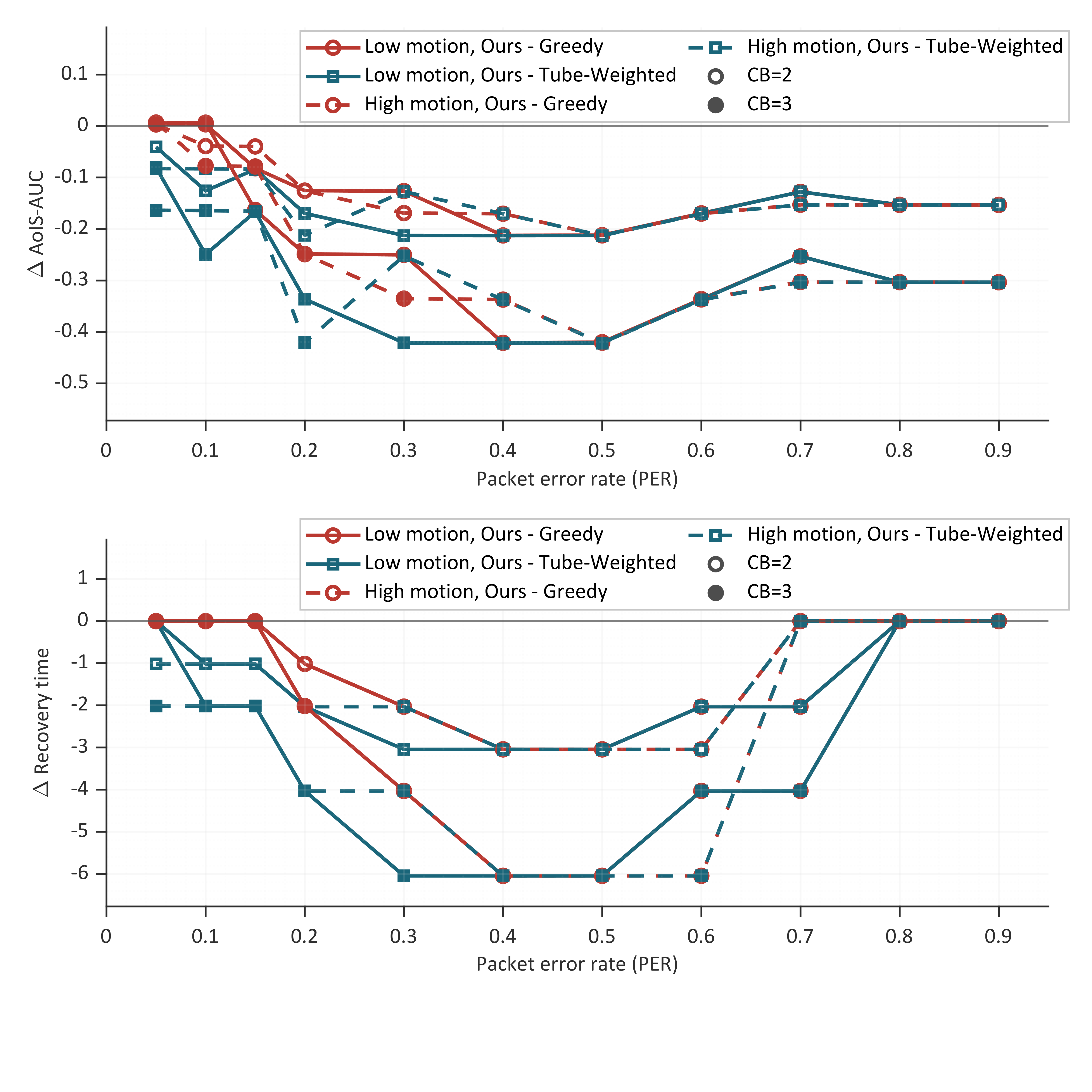}
  \caption{Motion-stratified AoIS-AUC and recovery-time gaps under PER sweeps.}
  \label{fig:motion_stratification}
\end{figure}

Fig.~\ref{fig:motion_stratification} shows that the improvement is not confined to a special subset of clips. The benefit appears in both motion strata, although its magnitude varies with motion level and baseline choice. Against greedy block requests, the high-motion stratum often shows an earlier onset or slightly larger gains in part of the moderate-PER range, which is consistent with the temporal-coherence argument. Against tube-weighted block requests, the low/high-motion gap is smaller and can reverse at some PER points, suggesting that part of the motion signal is already captured by tube-aware ordering. The appropriate conclusion is therefore not motion exclusivity, but motion-modulated gain with the overall primitive advantage unchanged. Together with the PER-sweep results, this indicates where package-native transport is most beneficial across channel and motion regimes.

\begin{figure}[!t]
  \centering
  \includegraphics[width=\linewidth,trim=0 10bp 0 0,clip]{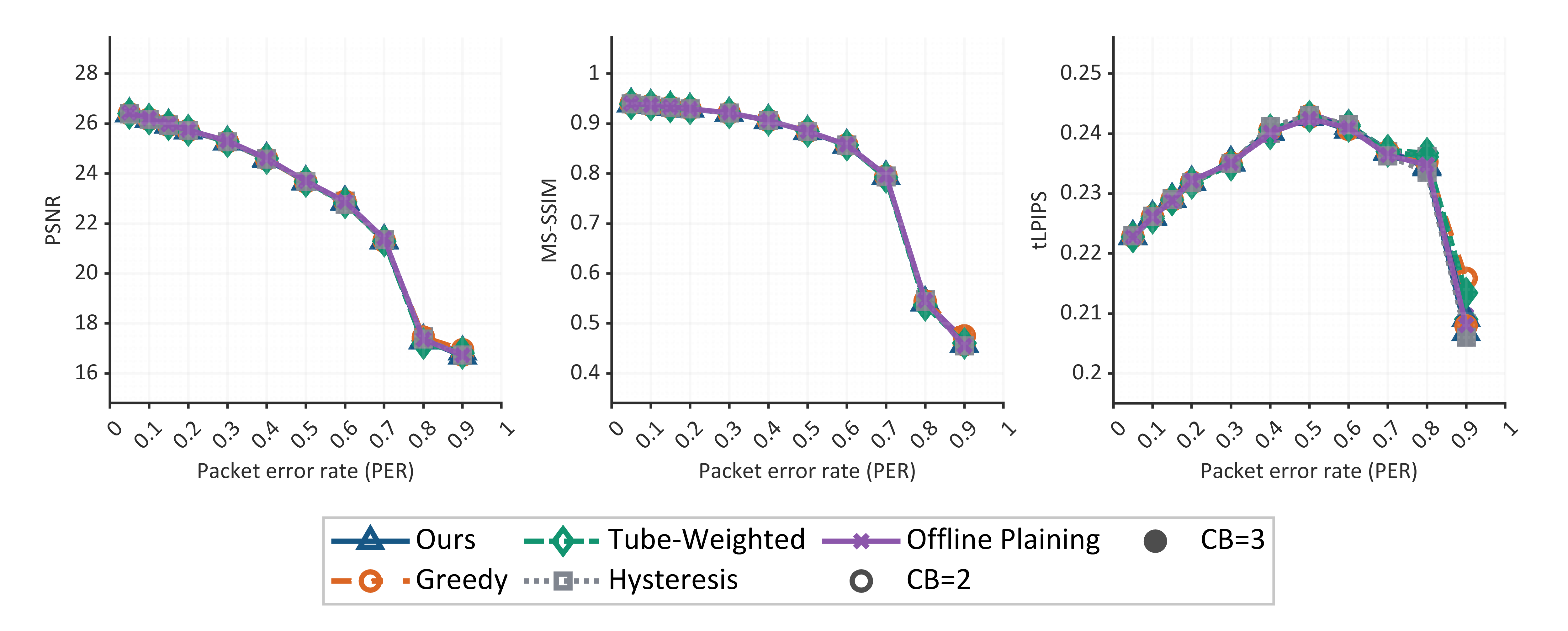}
  \caption{Final-quality curves under PER sweeps at $K=8$: PSNR, MS-SSIM, and tLPIPS.}
  \label{fig:quality_k8}
\end{figure}

Finally, Fig.~\ref{fig:quality_k8} checks whether the time-weighted gain is obtained at the expense of final reconstruction quality. The PSNR, MS-SSIM, and tLPIPS curves nearly overlap across most of the PER sweep, with no visually stable separation. The proposed primitive therefore does not show a systematic endpoint-quality tradeoff in either direction. Combined with Fig.~\ref{fig:per_sweep_k8} and Fig.~\ref{fig:recovery_delay}, this indicates that the tube-package design improves the primary AoIS-AUC objective mainly by accelerating stabilization rather than by materially changing the final reconstruction endpoint.

\vspace{-0.4em}
\subsection{Complexity Analysis}
\label{subsec:complexity_analysis}

The proposed method adds only lightweight request-side overhead on top of the shared generative backbone. The package catalog is built once per clip and then kept fixed throughout HARQ interaction. Online, each round scores and sorts the active package candidates before greedy budgeted selection, giving request-side complexity $O(P\log P)$ for a catalog of size $P$ under full sorting. By contrast, Greedy Block Requests and Tube-Weighted Block Requests rank over the active missing-block set, whose size is typically larger than the coarser package catalog.

The additional cost of the proposed method is confined to request-side bookkeeping rather than the dominant numerical kernel. The main runtime therefore remains the shared diffusion reconstruction, so the proposed primitive changes the interaction-side transport organization rather than the underlying generative reconstruction workload.

\vspace{-0.4em}
\section{Conclusion}

This paper studied receiver-driven semantic HARQ for generative video reconstruction under a budget-constrained AoIS-AUC objective and identified the retransmission primitive as an important design object. By introducing a tube-structured package-native primitive that makes temporally local packages channel-visible HARQ objects, the proposed design aligns the transport unit with the spatiotemporal persistence of video semantics. Under a controlled evaluation protocol with matched backbone, channel model, budgets, and metric, the proposed primitive improves the time-weighted recovery trajectory over competitive frame-block baselines in moderate-to-harsh regimes where interaction remains consequential, while the gain naturally shrinks in near-clean channels. The results indicate that this gain mainly comes from earlier stabilization through more coherent evidence delivery, without additional reconstruction cost and without systematic degradation in final reconstruction quality. Overall, retransmission primitive redesign provides a practical system-level complement to more capable reconstruction backbones in closed-loop generative semantic video systems.

\bibliographystyle{IEEEtran}
\bibliography{references}

\end{document}